\documentclass[aps,prl,10pt,twocolumn]{revtex4-1}
\usepackage{graphicx}
\usepackage{amsmath}
\usepackage{amssymb}
\usepackage{color}
\begin{document}

\title{Reciprocal-space-trajectory perspective on high harmonic generation in solids}

\author{Liang Li$^{1}$, Pengfei Lan$^{1}$}
\email{pengfeilan@hust.edu.cn}
\author{Xiaosong Zhu$^{1,2}$}
\email{zhuxiaosong@hust.edu.cn}
\author{Tengfei Huang$^{1}$}
\author{Qingbin Zhang$^{1}$}
\author{Manfred Lein$^2$}
\author{Peixiang Lu$^{1,3}$}
\email{lupeixiang@hust.edu.cn}

\affiliation{%
$^1$Wuhan National Laboratory for Optoelectronics and School of Physics, Huazhong University of Science and Technology, Wuhan 430074, China\\
$^2$Institut f\"{u}r Theoretische Physik, Leibniz Universit\"{a}t Hannover, D-30167 Hannover, Germany\\
$^3$Hubei Key Laboratory of Optical Information and  Pattern Recognition, Wuhan Institute of Technology, Wuhan 430205, China
}%

\date{\today}

\begin{abstract}
We revisit the mechanism of high harmonic generation (HHG) from solids by comparing HHG in laser fields with different ellipticities but a constant maximum amplitude. It is shown that the cutoff of HHG is strongly extended in a circularly polarized field. Moreover, the harmonic yield with large ellipticity is comparable to or even higher than that in the linearly polarized field. To understand the underlying physics, we develop a reciprocal-space-trajectory method, which explains HHG in solids by a trajectory ensemble from different ionization times and different initial states in the reciprocal space. We show that the cutoff extension is related to an additional pre-acceleration step prior to ionization, which has been overlooked in solids. By analyzing the trajectories and the time-frequency spectrogram, we show that the HHG in solids cannot be interpreted in terms of the classical re-collision picture alone. Instead, the radiation should be described by the electron-hole interband polarization, which leads to the unusual ellipticity dependence. We propose a new four-step model to understand the mechanism of HHG in solids.
\end{abstract}                         
\maketitle

High harmonic generation (HHG) from gas-phase atoms and molecules has opened up a new frontier in ultrafast science, where the generation of attosecond pulses \cite{Hentschel,Paul} and the measurement with attosecond temporal and Angstrom spatial resolutions become accessible \cite{Uiberacker2007,Baker,Blaga,Pertot}. In the very recent years, HHG has also been observed in solids \cite{Ghimire2011,TTLuu2015,Ghimire2016}, which makes it possible to extend the successful attosecond  metrology to solid systems.

HHG in two-dimensional (2D) nontrivially polarized laser fields in gases has attracted a great deal of attention in the past years, because it provides deeper insight into the mechanism of HHG \cite{Burnett1995,Budil1993} and also promises unprecedented applications such as generation of circularly polarized high harmonics \cite{circular1,circular2} and uncovering the ultrafast dynamics in atoms \cite{chiral}. Accordingly, the HHG in solids with the circularly or elliptically polarized laser fields has attracted close attention recently \cite{YouYS2016,Ndabashimiye,Dejean2017,Yoshikawa2017,Corkum2015}. However, because solid systems have much more complex structure and dynamical processes than gases, HHG in solids exhibits many distinct and unexplored features \cite{YouYS2016,Yoshikawa2017,Ndabashimiye}. Although several numerical models, such as numerically solving the time-dependent Sch\"{o}dinger equation (TDSE) \cite{MW2015,Bian2017,Takuya2017,Lewenstein2017}, the semiconductor Bloch equation (SBE) \cite{Koch2003,Golde2008,Vampa2014,TTLuu2016,Jiangsc2018}, and time-dependent density functional theory (TDDFT) \cite{Runge1984,Leeuwen1998} provide good descriptions of HHG, the underlying mechanisms are buried in the wave functions. Following the principle of HHG in gases \cite{Corkum1993}, a generalized three-step model \cite{Vampa2014,Vampa2015,Jiangsc2017} has also been proposed for solids. In brief, near the peak of the laser field, the electron tunnels vertically from the valence band (VB) to the conduction band (CB). Then, the electron is accelerated and it recombines some time later with emission of a high harmonic photon. This model provides a useful and intuitive explanation of HHG in solids, but fails to give a satisfactory quantitative description. A quantitative model in terms of electron trajectories is still not established. Therefore, especially for 2D laser fields, a clear understanding has not been reached so far.

In this Letter, we investigate the HHG in solids driven by circularly and elliptically polarized laser fields. Our results indicate that the HHG spectra show a clear cutoff extension in a circularly polarized field. Moreover, by fixing the amplitude of electric field, the harmonic yield with large ellipticity is comparable to or even higher than that in the linearly polarized field. To explain these phenomena, we develop a reciprocal-space-trajectory (RST) method based on accelerated Bloch states, in which the HHG can be described in terms of the quantum path integral of the trajectory ensemble. This method fully includes quantum interference effects and allows us to reproduce the HHG spectrum. Based on the analysis of electron trajectories, the cutoff extension is well explained by the pre-acceleration of the electrons before being excited to the CB. Moreover, the abnormal ellipticity dependence of the HHG yield is better explained by the emission associated with the polarization of the electron-hole pairs rather than the previously believed classical re-collision \cite{YouYS2016}. Our study suggests a new four-step picture of HHG in solids.

\begin{figure}[!t]
  \includegraphics[width=9cm]{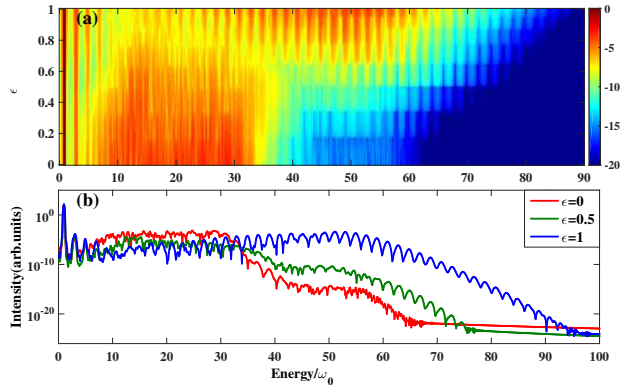}
  \caption{(a) High harmonic spectra for different ellipticities $\varepsilon = F_{y0}/F_{x0}$ A logarithmic color scale is used. (b) High harmonic spectra for $\varepsilon=$ 0, 0.5 and 1.}\label{fig1}
\end{figure}

Figure \ref{fig1} shows the high harmonic spectra of ZnO driven by laser pulses with different ellipticities.  We use the same band structure as in Ref. \cite{Vampa2014,Goano2007}. The orientation of the reciprocal lattice of ZnO is chosen as $\hat{\textbf{x}}||\Gamma-M$, $\hat{\textbf{y}}||\Gamma-K$ and $\hat{\textbf{z}}||\Gamma-A$. The harmonic spectra are obtained by numerically solving the SBE. The dephasing effect does not change the structure of the harmonic spectra as shown in Ref. \cite{Vampa2014} and is neglected in this work. For computational convenience, we perform two-dimensional (2D) calculations (i.e., $k_z=0$ in the reciprocal space), which can well reproduce the features of HHG in 3D simulations (see the Supplemental Material \cite{SM}). The frequency of the driving field is $\omega = 0.014$ atomic units (a.u.). The laser pulse is polarized in the x-y plane, and the ellipticity is varied by keeping the laser amplitude in the major axis constant ($F_{x0}=0.004$ a.u.) and changing the amplitude in the minor axis ($F_{y0}$) from 0 to $0.004$ a.u. The electric field oscillates under a sine squared envelope with a duration of 10 optical cycles. As shown in Fig. \ref{fig1}, the harmonic spectra show an obvious cutoff extension in a circularly polarized field. With decreasing ellipticity, the spectrum changes into a two-plateau structure. This two-plateau structure is different from that contributed by the multiple conduction bands \cite{MW2015}, because the spectral range is within the energy gap between the states of the VB and the first CB. Moreover, it is striking that, the harmonic yield in the circularly and elliptically polarized fields is comparable to or even higher than that in the linearly polarized field. These features are very different from those in gaseous media, where the harmonic yield decreases dramatically with the increase of the ellipticity and almost no high harmonic is generated at large ellipticities \cite{MM2012}. Apparently, the ellipticity dependence here is different from generally recognized ones.

The purpose of this work is to give a clear physical picture and bring new physical insight of the HHG in solids. To this end, we develop an intuitive RST method to understand the above numerical results. The interaction between the harmonics field and the electrons is neglected and the dipole approximation is applied. The instantaneous eigenstate $|\varphi(t)\rangle$ of Hamiltonian $\hat{H}(t)$ with eigenenergy $\tilde{E}(t)$ in velocity-gauge satisfies:
\begin{align}\label{Hamiltonian}
  \hat{H}(t)|\varphi(t)\rangle = (\frac{[\hat{\textbf{p}}+\textbf{A}(t)]^{2}}{2}+U(\hat{\textbf{r}}))|\varphi(t)\rangle = \tilde{E}(t)|\varphi(t)\rangle,
\end{align}
where the periodicity of the Hamiltonian is maintained. $\hat{\mathbf{p}}$ is the momentum operator, $U(\hat{\textbf{r}})$ is the crystal potential, and $\mathbf{A}(t)$ is the vector potential of the laser field. The solutions of Eq. (\ref{Hamiltonian}) satisfy the Bloch theorem and the Born-von K\'{a}rm\'{a}n boundary conditions \cite{Krieger1986}, thus
\begin{align}\label{eigenstates}
  \langle\textbf{r}|\varphi_{m,\textbf{k}_{0}}(t)\rangle &= e^{-i\textbf{A}(t)\cdot\textbf{r}}\phi_{m,\textbf{k}(t)}(\textbf{r}), \nonumber \\
  \tilde{E}_{m,\textbf{k}_{0}}(t) &= E_{m}(\textbf{k}(t)),
\end{align}
where $m$ is the band index and $\textbf{k}(t) = \textbf{k}_{0}+\textbf{A}(t)$ is the crystal momentum. $\textbf{k}(t)$ obeys the acceleration theorem $d\textbf{k}(t)/dt = -\textbf{F}(t) = d\textbf{A}/dt$. $\textbf{F}(t)$ is the electric field. $\textbf{k}_0$ is the crystal momentum of the electron at the initial time $t_0$. $\phi_{m,\textbf{k}}(\textbf{r})$ is the Bloch state and $E_{m,\textbf{k}}$ is the energy of the band $m$ and crystal momentum $\textbf{k}$. The states $|\varphi_{m,\textbf{k}_{0}}(t)\rangle$ are the accelerated Bloch states \cite{Yakoelev2016}.

\begin{figure}[!t]
  \includegraphics[width=6cm]{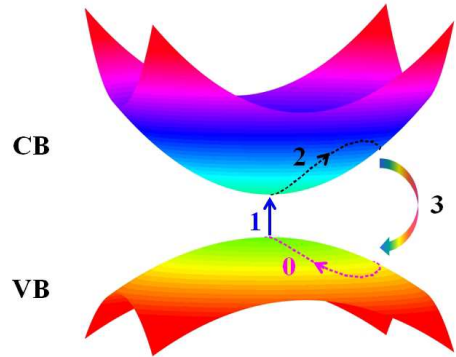}
  \caption{The sketch of electron trajectories in reciprocal space. Only one schematic trajectory is shown. The red dashed line marks the pre-acceleration in VB, and the black dashed line marks the acceleration in CB.}\label{fig2}
\end{figure}

Based on the accelerated Bloch states, the evolution of the electron wave packet can be described as \cite{Yakoelev2016}
\begin{align}\label{wave function}
  |\Psi_{\textbf{k}_{0}}(t)\rangle = \sum_{m}\alpha_{m,\textbf{k}_{0}}(t)e^{-i\int_{t_{0}}^{t}d\tau E_{m}(\textbf{k}(\tau))}|\varphi_{m,\textbf{k}_{0}}(t)\rangle,
\end{align}
where $\alpha_{m,\textbf{k}_{0}}(t)$ is the complex amplitude for the contribution in band $m$. The corresponding wave vector $\textbf{k}(t)$ of the Bloch state constitutes a trajectory in reciprocal-space. In the framework of RST, one can obtain a trajectory perspective on HHG as illustrated in Fig. \ref{fig2}. For simplicity, we consider only two bands i.e., $m=c,v$ for the CB and VB, which has been shown to work well in previous works \cite{Vampa2014,Vampa2015,Osika2017}. An electron initially located at $\textbf{k}_{0}$ in the VB undergoes the following steps, where ``ionization'' is to be understood as excitation to the CB:

(0) Before ionization, the electron is accelerated in the VB. This is different from the picture in gases where the electron is localized in the ground state prior to tunnel ionization. It is also in contrast to the usual assumption in the previous three-step model for HHG in solids, where the electron is initially located at the top of the VB and vertically promoted into the CB at the same crystal momentum \cite{MW2015,Bian2017}. We refer to this acceleration process prior to ionization as \textit{pre-acceleration}.

(1) The electron can be excited from the VB to the CB at time $t'$ with certain probability. When the depletion is neglected, one obtains the complex amplitude of the ionization rate $\chi_{\textbf{k}_{0}}(t') = e^{iS_{c,\textbf{k}_{0}}(t',t_0)}\Omega_{cv}(t')e^{-iS_{v,\textbf{k}_{0}}(t',t_0)}$ \cite{SM}. $\Omega_{cv}(t) = \vec{\xi}_{cv}(\textbf{k}(t))\cdot \textbf{F}(t)$ involves the transition matrix elements $\vec{\xi}_{cv}(\textbf{k}) = \langle c,\textbf{k}|\nabla_{\textbf{k}}|v,\textbf{k}\rangle_{cell}=\frac{1}{V}\int_V d^3r u^*_{c,\textbf{k}}(\textbf{r})\nabla_{\textbf{k}}u_{v,\textbf{k}}(\textbf{r})$, where the integration is performed over a unit cell and $u_{m,\textbf{k}}$ is the periodic part of Bloch state. $\vec{\xi}_{cv}(\textbf{k})$ can be obtained from the ab initio calculations \cite{Goano2007} and can be simplified with appropriate approximations following \cite{Vampa2014}. The relative phase includes two parts: $S_{v,\textbf{k}_{0}}(t',t_{0}) = \int_{t_{0}}^{t'}d\tau E_{v}(\textbf{k}(\tau))$ and $S_{c,\textbf{k}_{0}}(t',t_{0}) = \int_{t_{0}}^{t'}d\tau E_{c}(\textbf{k}(\tau))$ accumulated in the VB and CB, respectively. Note that the trajectories of different ionization times are coherent with each other and the ionization is dominant at the peak of the electric field.

(2) After ionization, the electron is accelerated by the external field and the trajectory in reciprocal space $\textbf{k}(t)$ satisfies the acceleration theorem. The motions of the electron and the hole are different in the real-space, while their trajectories in the reciprocal-space concerned. During the acceleration process, the instantaneous electron energy is $E_{c}(\textbf{k}(t))$ and $E_{v}(\textbf{k}(t))$ in the CB and VB, respectively. Their energy difference is denoted as $\Delta E(\textbf{k}(t))=E_{c}(\textbf{k}(t))-E_{v}(\textbf{k}(t))$.

(3) During the oscillation, two mechanisms contribute to the high harmonic emission. One is the intraband current induced by the charge transfer inside the VB and the CB, respectively. The other one is the interband current contributed by the coherence between the electron in the CV and VB. The induced intraband and interband currents are expressed as:
\begin{align}\label{currents}
  &\textbf{j}_{mm,\textbf{k}_{0}}(t) = \textbf{p}_{mm}(\textbf{k}(t)), \nonumber \\
  \textbf{j}_{cv,\textbf{k}_{0}}(t) &= e^{iS_{c,\textbf{k}_{0}}(t,t_{0})}\textbf{p}_{cv}(\textbf{k}(t))e^{-iS_{v,\textbf{k}_{0}}(t,t_{0})},
\end{align}
where $\textbf{p}_{mm}(\textbf{k}) = \langle m,\textbf{k}|\hat{\textbf{p}}|m,\textbf{k}\rangle = \nabla_{\textbf{k}}E_{m}(\textbf{k})$, $\textbf{p}_{cv}(\textbf{k}) = \langle c,\textbf{k}|\hat{\textbf{p}}|v,\textbf{k}\rangle$.

In general semiconductors, the VB is fully occupied. The trajectories from different ionization times and initial states form a trajectory ensemble. In this case, the total currents can be expressed as $J(t) = J^\textrm{intra}(t)+J^\textrm{inter}(t)$ with \cite{SM}
\begin{align}\label{Scurrents}
  \textbf{J}^\textrm{intra}(t) &= \sum_{\textbf{k}_{0}\in \textrm{BZ}}|\int_{t_{0}}^{t}dt' \alpha_{c,\textbf{k}_{0}}(t,t')|^2[\textbf{j}_{cc,\textbf{k}_{0}}(t)-\textbf{j}_{vv,\textbf{k}_{0}}(t)],  \nonumber \\
  \textbf{J}^\textrm{inter}(t) &= \sum_{\textbf{k}_{0}\in \textrm{BZ}}\int_{t_{0}}^{t}dt'\alpha^{*}_{c,\textbf{k}_{0}}(t,t')\textbf{j}_{cv,\textbf{k}_{0}}(t)+c.c.,
\end{align}
where $\alpha_{c,k_0}(t,t')=\chi_{\textbf{k}_{0}}(t')\theta(t,t')$ describes the weight of the electron trajectories initially located at $\textbf{k}_0$ and ionized at time $t'$. $\theta(t,t')$ is Heaviside function with $\theta(t,t')=0$ for $t< t'$ and $\theta(t,t')=1$ for $t\geq t'$. The HHG spectrum can be obtained by Fourier transformation of the currents. Within the parameters used in this work, the harmonic yield above the minimum band gap is dominated by the interband current (see the Supplementary Material \cite{SM}). Hence, hereafter we concentrate on the HHG contributed by interband current.


\begin{figure}[!t]
  \includegraphics[width=9cm]{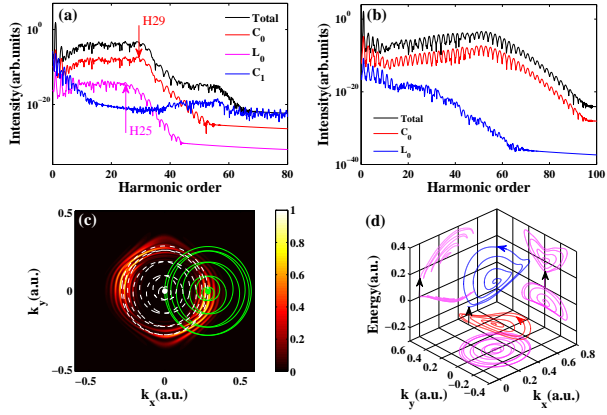}
  \caption{Harmonic spectra obtained with different channels in (a) linear and (b) circular fields. For clarity, the spectrum are shifted by constant factors $10^{-5}$ for $C_{0}$ and $C_{1}$, and $10^{-10}$ for $L_{0}$. The laser parameters are the same as that in Fig. 1. (c) Normalized $\textbf{k}_0$-dependent harmonic yield contributed by the trajectories of electrons initially located at different positions in reciprocal space. The green solid line and white dashed line mark representative trajectories of $C_{0}$ and $L_{0}$, respectively. (d) The representative trajectory of the channel $C_{0}$ in $k_x-k_y-\text{energy}$ space.}\label{fig3}
\end{figure}

To validate the above model, we calculate the HHG spectra from ZnO under the same conditions as in Fig. \ref{fig1}. The HHG spectra obtained with RST method agree well with those from the SBE simulations (see the Supplementary Material \cite{SM}). More importantly, the RST model enables us to anatomize HHG into contributions from different trajectories and understand how they build up the structure of the total harmonic spectra.
We first discuss the HHG in a linearly polarized field. For the RST method, we perform a 2D calculation by considering trajectories with initial states in the plane $k_{z}=0$. By analyzing the HHG contributed by different trajectories, one can find the link between the HHG and the intuitive trajectories. We consider the electron trajectories ionized at the top of the VB (including $1\%$ area of the BZ), denoted as channel $C_{0}$. According to Eqs. (\ref{currents}) and (\ref{Scurrents}), the high harmonic contributed by the channel $C_{0}$ can be obtained:  $Y_{C_{0}} = \text{FFT}\{\sum_{\textbf{k}_{0}+\textbf{A}(t')\in C_{0}}[(|\int dt'\alpha_{c,\textbf{k}_{0}}(t,t')|^2[\textbf{j}_{cc,\textbf{k}_{0}}(t)-\textbf{j}_{vv,\textbf{k}_{0}}(t)])+(\int dt'\alpha^{*}_{c,\textbf{k}_{0}}(t,t')\textbf{j}_{cv,\textbf{k}_{0}}(t)+c.c.)]\}$. As a comparison, following Refs. \cite{MW2015,Bian2017}, we also consider the electron trajectories initially located at the top of the VB (including $1\%$ area of the BZ), which we denote as channel $L_{0}$. The harmonic spectrum with the full trajectory ensemble is also presented in Fig. \ref{fig3}(a). One can see that, although the spectrum obtained by the channel $L_{0}$ shows a plateau, the cutoff energy (25$\omega$) is smaller than the first plateau cutoff of the total spectrum (29$\omega$). In contrast, the spectrum obtained by the channel $C_{0}$ can well reproduce the main structure and the cutoff position of the total spectrum. The channel $C_{0}$ contains the trajectories from different initial locations satisfying $\textbf{k}_{0}=-\textbf{A}(t_{i})$ ($t_{i}$ is the ionization time). Due to the pre-acceleration step, these electrons are driven to reach the top of the VB at the instant of ionization. Since the energy gap at the top of the VB is smallest, these electrons contribute predominantly to the total HHG. The farther the electrons are initially located away from the top of VB, the larger $\textrm{max}\{\Delta E[-\textbf{A}(t_{i})+\textbf{A}(t)]\}$ can be gained. Therefore, a larger cutoff energy can be obtained by channel $C_{0}$, and the first plateau in the total spectrum is well reproduced. In contrast, the trajectories of $L_{0}$ are overlapped with each other and only gain the maximum energy of $\textrm{max}\{\Delta E[\textbf{A}(t)]\}$, which is smaller than $\textrm{max}\{\Delta E[-\textbf{A}(t_{i})+\textbf{A}(t)]\}$.

Next, we investigate the much weaker harmonic signal in the second plateau (from the 40th to 60th harmonic orders). We plot the HHG spectrum contributed by the electron trajectories ionized far away from the top of VB (denoted as $C_{1}$). As shown in Fig. \ref{fig3}(a), the electron trajectories of $C_{1}$ channel cover much higher energy region than that of $C_{0}$ and contribute to the harmonics in the second plateau. Since the ionization of $C_{1}$ occurs with a large band gap, the HHG yield is much lower than that of channel $C_{0}$.

The above results reveal an important role of the pre-acceleration, allowing the electrons located far away from the top of VB to be accelerated to the top of VB and efficiently contribute to HHG. Note that the vector potential (i.e. $|\textbf{A}(t_{i})|$) is near zero when the electric field reaches its peaks in a linearly polarized field, therefore the difference between $\textrm{max}\{\Delta E[-\textbf{A}(t_{i})+\textbf{A}(t)]\}$ and $\textrm{max}\{\Delta E[\textbf{A}(t)]\}$ is not so large. This is a possible reason why the importance of pre-acceleration has not been noticed before.

The effect of pre-acceleration will be more obvious in elliptically or circularly polarized fields. As above, we consider the spectra obtained by the channels $C_{0}$ and $L_{0}$ and compare them with the total spectrum. As shown in Fig. \ref{fig3}(b), the spectrum obtained from the channel $C_{0}$ reproduces the total spectrum well. In contrast, the channel $L_{0}$ shows a much lower cutoff energy. To obtain an intuitive understanding of this result, we analyze the trajectories and the harmonic yield contributed by the trajectories of electrons located at different initial momenta $\textbf{k}_0$. As a quantitative measure of the yield from initial momentum $\textbf{k}_0$, we calculate
\begin{align}\label{Averagingyield}
\overline{Y}_{\textbf{k}_{0}} = \frac{\int_{\min[\Delta E(\textbf{k}_{0}+\textbf{A}(t))]}^{\max[\Delta E(\textbf{k}_{0}+\textbf{A}(t))]}Y_{\textbf{k}_{0}}(\omega)d\omega}{\max[\Delta E(\textbf{k}_{0}+\textbf{A}(t))]-\min[\Delta E(\textbf{k}_{0}+\textbf{A}(t))]},
\end{align}
where $Y_{\textbf{k}_{0}}(\omega)$ is the HHG yield contributed by electrons trajectories from $\textbf{k}_0$. As shown in Fig. \ref{fig3}(c), the dominant distribution of the $\textbf{k}_0$-dependent harmonic yield shows a donut structure at about $|\textbf{k}_0|=A_{0}=F_{0}/\omega$, where $A_{0}$ and $F_{0}$ are the amplitude of the vector potential and electric field. This shape is familiar from the photonionization of gases in circularly polarized fields \cite{Bian2011}, where electrons are born at non-zero vector potential. The exact shape of the dominant distribution will be related to the band structure of the system. In this work, the band structure is approximately rotational symmetric around the $k_{z}$ axis like those in atoms, so nearly round structure is revealed. The solid green line in Fig. \ref{fig3}(c) marks one representative trajectory of channel $C_{0}$, which is initially located at $|\textbf{k}_{0}|=F_{0}/\omega$. The trajectory is also presented in the $k_{x}$-$k_{y}$-$\text{energy}$ space, see Fig. \ref{fig3}(d). The red line marks the pre-acceleration in VB. As already discussed in Fig. \ref{fig3}(a), electron trajectories initially located farther away from the top of VB contribute to a higher energy region of HHG than the trajectories initially located close to the top of the VB. Consequently, a harmonic spectrum with both broad spectral range is obtained.

\begin{figure}[!t]
	\includegraphics[width=8cm]{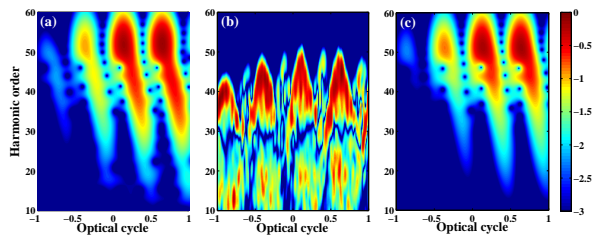}
	\caption{The time-frequency spectrogram from (a) numerically solving SBE, (b) re-collision trajectories, and (c) polarization of electron-hole pairs. The maximum values are normalized to 1 for clarity.  A logarithmic color scale is used.}\label{fig4}
\end{figure}

Finally, to check the general validity of the re-collision picture, we calculate the time-frequency spectrogram of HHG in the circularly polarized field by accumulating the trajectories that re-collide with the parent hole or the neighbor holes (see \cite{SM} for details). The result is shown in Fig. \ref{fig4}(b). One can see two emissions in each half optical cycle. It is attributed to the nearly four-fold symmetry of the crystal structure \cite{SM}. For comparison, Fig. \ref{fig4}(a) shows the time-frequency spectrogram obtained by numerically solving the SBE. Different from Fig. \ref{fig4}(b), the SBE simulations show only one emission every half cycle and each emission has a negative slope versus time. These differences indicate the failure of the re-collision picture in solid HHG with circularly polarized fields. Note that, for gas-phase atoms, the ground state is tightly localized around the core. The harmonic emission is quenched in the circular field since electrons and holes miss each other. But the situation is very different in solids. The hole states are Bloch waves and can be delocalized. There is still substantial overlap between the sub-waves of the electron-hole pair contributing to the interband current, even though the electron does not undergo classical re-collision with either the parent or the neighbor holes. In this case, the harmonic emission is better understood in terms of the polarization of the electron-hole pairs rather than the re-collision scheme. The time-frequency distribution of interband emission in terms of the RST is obtained as
\begin{align}\label{TF}
&F(t,\omega') \nonumber \\
&= |\sum_{C_{0}}\int dt'\pi(t',t_{i})e^{i\omega't'}e^{-\frac{(t'-t)^{2}}{(\gamma)^{2}}}|^2
\end{align}
$\pi(t,t_{i}) = \chi^{*}_{\textbf{k}_{0}}(t_{i})e^{iS_{c,\textbf{k}_{0}}(t,t_{0})-iS_{v,\textbf{k}_{0}}(t,t_{0})}\textbf{p}_{cv}$ is the time-dependent polarization for a trajectory ionizing at time $t_{i}$. The summation is over the channel $C_{0}$. Here, we consider the interference of the trajectories in a time window with width $\gamma = T_{0}/6$. As shown in Fig. \ref{fig4}(c), both the emission time and emission energy agree well with the SBE model.

Before concluding, we address the issue how the electron-hole polarization is related to the re-collision mechanism in the linearly polarized field. In this case, the overlap between the sub-waves of electron-hole pairs is largest for the re-colliding trajectories. Then the polarization of electron-hole pairs is dominated by the re-collision trajectories. This explains why semiclassical trajectories can explain the time-frequency features of solid HHG in the linearly polarized field \cite{Vampa2015,Hansen,Garg}.

In conclusion, we have investigated HHG from ZnO driven by laser fields with various ellipticities. We find that the harmonic cutoff is substantially extended in the circularly polarized field and the harmonic yield is comparable to or higher than that in the linearly polarized field. To explain this phenomenon, we propose a new four-step model for the HHG in solid. Correspondingly, we developed an RST method, which relates the HHG response to the microscopic dynamics of electrons. It is shown that a previously overlooked pre-acceleration step plays an important role, which selects electron trajectories starting from locations other than the top of the VB to predominantly contribute to HHG. Moreover, the re-collision picture fails in the circularly polarized field, where the HHG is better explained by the polarization of electron-hole pairs. The RST model brings a clear physical picture of the HHG in solids. It will become a useful tool to understand the HHG in solid with more complex structure or in more complicated driving fields.

\begin{acknowledgments}
We gratefully acknowledge G. Vampa for valuable discussions.
This work was supported by the National Natural Science Foundation of China
under Grants Nos. 11874165, 11627809, and 11774109, and the Alexander von Humboldt Foundation.
\end{acknowledgments}

\end{document}